\documentclass[12pt]{article}
\usepackage{amsmath, amssymb, latexsym, graphicx}
\begin{document}
\newcommand{\di}{\displaystyle}
\newcommand{\reff}[1]{(\ref{eq:#1})}
\newcommand{\labl}[1]{\label{eq:#1}}
\newcommand{\Rot}{{\pmb{\nabla}}\times}
\newcommand{\Grad}{{\pmb{\nabla}}}
\newcommand{\Div}{{\pmb{\nabla}}\cdot}
\newcommand{\Gn}{\ensuremath{{\bf G}^{(0)}}}
\newcommand{\Gnc}[1]{\ensuremath{G^{(0)}_{#1}}}
\newcommand{\En}{\ensuremath{{\bf E}^{(0)}}}
\newcommand{\Enc}[1]{\ensuremath{E^{(0)}_{#1}}}\newcommand{\Gnp}{\ensuremath{{\bf G}'^{(0)}}}
\newcommand{\Anp}{\ensuremath{{\bf A}'^{(0)}}}
\newcommand{\Fn}{\ensuremath{{\bf F}^{(0)}}}
\newcommand{\Fnp}{\ensuremath{{\bf F}'^{(0)}}}
\newcommand{\Fnpc}[1]{\ensuremath{F'^{(0)}_{#1}}}
\newcommand{\Bnp}{\ensuremath{{\bf B}'^{(0)}}}

\title{On the Nonlocality of the Coulomb Gauge External Field Problem}
\date{}
\maketitle

\vspace{-15mm}

\begin{center}
\author{P\'{e}ter Hrask\'{o}\\
University of P\'{e}cs, Hungary, H-7622 P\'{e}cs, Vasv\'{a}ri P\'{a}l utca 4\\
Email: peter@hrasko.com}
\end{center}

\bigskip

\begin{abstract}
The apparent nonlocality of the Coulomb gauge external field problem in electrodynamics is illustrated with an example in which nonlocality is especially striking. Explanation of this apparent nonlocal behaviour based on a purely local picture is given. A gauge invariant decomposition of the Lorentz-force into two terms with clear physical meanings is pointed out. Based on this decomposition  derivation of the Aharonov-Bohm effect in terms of field strengths alone is given.
\end{abstract}

\section{Az introductory example}

Consider a rigid body with a given charge distribution $\rho({\bf x})$ on it but of total charge equal to zero. If it is placed into a homogeneous magnetic field $\bf B = \Rot{\bf A}$ then in the combined electric field $\bf E$ of the body and the external magnetic field a certain amount $\bf N$ of angular momentum will arise. The simplest way to demonstrate this is to slowly remove the magnetic field and to calculate the torque
\begin{equation}
{\bf K}(t) = \int d^3x\,[{\bf x}, \rho{\bf E}'(t)]\labl{1a}
\end{equation}
acting on the body due to the electric field ${\bf E}'(t)$ induced by the changing $\bf B(t)$. The total amount of angular momentum imparted to the body is equal to the angular momentum $\bf N$.

In Coulomb gauge the vector potential is chosen so as to make its divergence to vanish and the scalar potential $\phi$ is equal to the Coulomb field of the charge density $\rho({\bf x})$. For a homogeneous field we have
\begin{equation}
{\bf A} = \frac{1}{2}[{\bf B},\,{\bf x}]\qquad (\Div {\bf A} = 0).\labl{2a}
\end{equation}
The induced electric field is determined by the equation
\[{\bf E}'(t) = -\frac{1}{c}\dot{\bf A}(t),\qquad {\bf A}(0) = {\bf A}, \qquad {\bf A}(\infty ) = 0.\]
Then \reff{1a} can be written as 
\[{\bf K}(t) = -\frac{d}{dt}\int d^3x\,[{\bf x}, \frac{\rho}{c}{\bf A}(t)].\]
Since $\dot{\bf N} = {\bf K}$ the total amount of angular momentum received by the body is equal to
\begin{equation}
{\bf N} = +\int d^3x\,[{\bf x}, \frac{\rho}{c}{\bf A}].\labl{1b}
\end{equation}

The angular momentum stored in the field can, therefore, be written as
\[{\bf N} = \frac{1}{2c}\int d^3x\,\rho ({\bf x})\Bigl [{\bf x},\,[{\bf B},\,{\bf x}]\Bigr ].\]
Since
\[\Bigl [{\bf x},\,[{\bf B},\,{\bf x}]\Bigr ] = -\frac{1}{3}\bigl (3({\bf x}\cdot {\bf B}){\bf x} - r^2{\bf B}\bigr ) + \frac{2}{3}r^2{\bf B},\]
where $r^2 \equiv {\bf x}^2$, the Cartesian components of $\bf N$ are
\begin{equation}
N_\alpha = -\frac{1}{6c}D_{\alpha\beta}B_\beta + \frac{1}{3c}\langle\rho r^2\rangle B_\alpha.\labl{3a}
\end{equation}
Here
\[D_{\alpha\beta} = \int d^3x\,\rho ({\bf x})(3x_\alpha x_\beta - r^2\delta_{\alpha\beta})\]
is the quadrupole momentum tensor of the body and
\[\langle\rho r^2\rangle = \int d^3x\,\rho ({\bf x})r^2\]
is its scalar part.

Simple as it seems our example reveals its rather peculiar feature when one tries to recalculate $\bf N$ starting from the equation 
\begin{equation}
{\bf N} = \frac{1}{4\pi c}\int\,d^3x \Bigl [{\bf x},\,[{\bf E},\,{\bf B}]\Bigr ]\labl{3b}
\end{equation}
which defines field angular momentum in terms of field strengths. To this end we replace the fields in \reff{3b} by their respective potentials  ${\bf B} = \Rot {\bf A}$ and ${\bf E} = - \Grad\phi$ and perform partial integrations within a sphere of arbitrary large radius $R$. Though, neglecting surface terms, we arrive at the expression \reff{3a} again, {\em the contribution of the surface terms turns out to be of finite magnitude independent of} $R$ (see Appendix):
\begin{equation}
\textstyle{\mathrm{Surface\; Terms}} = \di\frac{1}{10c}D_{\alpha\beta}B_\beta.\labl{4b}
\end{equation}
Hence we obtain that, contrary to \reff{3a}, within any sphere surroundig the body the amount of angular momentum is equal to
\begin{equation}
N^{(R)}_\alpha = -\frac{1}{15c}D_{\alpha\beta}B_\beta + \frac{1}{3c}\langle\rho r^2\rangle B_\alpha.\labl{4a}
\end{equation}
Moreover, this momentum is concentrated within the volume of the body since, owing to the independence of \reff{4b} of $R$, in any spherical shell outside the body the value of the angular momentum is cancelled to zero.

The correct conclusion from this apparent contradiction is that the notion of the homogeneous magnetic field extending to infinity is in general an unacceptable abstraction. In local problems as e.g. in Zeeman effect,
or even in the derivation of eq. \reff{3a}, such a concept is a perfectly suitable idealization but when total field momentum or angular momentum are to be calculated the sources of $\bf B$ should be taken into consideration explicitely.

Their simple realization is an infinitely long ideal straight solenoid (an idealization too but this time a harmless one) of very large circular cross section of radius $R_s$ with our body situated on its axis (see Figure). Then an amount of angular momentum given by \reff{4a} is concentrated within the volume of the body while the remaining part of it, the negative of \reff{4b}, is found within the solenoid above and below the sphere of radius $R_s$, arbitrarily far away from the charged body (see Appendix). The sum total is then given by \reff{3a} as expected.

\begin{figure}[h]
\centering%
\includegraphics[width=40mm]{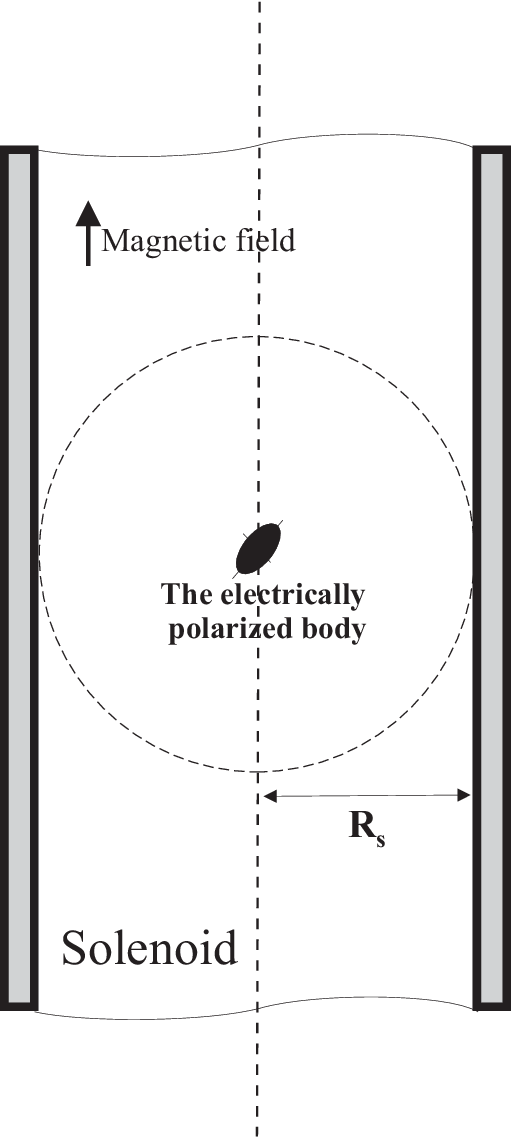}
\end{figure}

\section{The problem of apparent nonlocality}

Assume now that the body we are considering is capable to rotate around a fixed point of it. Then its Lagrangian is
\[L = \frac{1}{2}I_{\alpha\beta}\Omega_\alpha\Omega_\beta + \frac{1}{c}\int d^3x\,\rho ({\bf x}){\bf A}({\bf x})\cdot{\bf V}({\bf x}),\]
in which $I_{\alpha\beta}$ and ${\bf\Omega}$ are the inertia tensor and the angular velocity and ${\bf V} = [{\bf\Omega},\,{\bf x}]$. 

In Coulomb gauge, using \reff{2a} for the vector potential and the relation ${\bf A}[{\bf\Omega},\,{\bf x}] = {\bf\Omega}[{\bf x},\,{\bf A}]$, the interaction term becomes a linear function of the field angular momentum $\bf N$ of \reff{1b}:
\[L = \frac{1}{2}I_{\alpha\beta}\Omega_\alpha\Omega_\beta + {\bf N}\cdot{\bf\Omega}.\]
In the Hamiltonian $H = {\bf J}\cdot{\bf\Omega} - L$ 
\[J_\alpha = \frac{\partial L}{\partial\Omega_\alpha} = I_{\alpha\beta}\Omega_\beta + N_\alpha.\]
From this
\[\Omega_\alpha = (I^{-1})_{\alpha\beta}(J_\beta - N_\beta ).\]
Substituting this into $H$ we obtain after some rearragements
\begin{equation}
H = \frac{1}{2}(I^{-1})_{\alpha\beta}(J_\alpha - N_\alpha ).(J_\beta - N_\beta ).\labl{6b}
\end{equation}
The canonical variables in this Hamiltonian are the Euler angles which determine the orientation of our rotator and their conjugate momenta. We do not recall the explicit dependence of $\bf J$ and $\Omega_{\alpha\beta}$ on these variables. According to \reff{3a} $\bf N$ depends on the Euler angles through the quadrupole momentum tensor $D_{\alpha\beta}$.

The Hamiltonian is the energy of the rotator. The Lorentz-force by which $\bf B$ acts on the rotator does not perform any work on it and so $H$ is equal to its kinetic energy. Hence, the difference $({\bf J} - {\bf N})$ must be equal to the angular momentum $\bf R$ of its rotation. The canonical angular momentum $\bf J$ can, therefore, be interpreted as the sum of the rotational angular momentum and of the angular momentum stored in the combined external magnetic field and electric field of the body  {\em at rest}. This circumstance expresses the most characteristic feature of the Coulomb gauge: the potential $\phi$ of a time dependent charge distribution $\rho ({\bf x}, t)$ is equal to its {\em instantaneous} Coulomb potential [1]

In the previous section we have found that, depending on $D_{\alpha\beta}$ (i.e. on the orientation of the body), a well defined fraction of the field angular momentum is found at a distance arbitrarily far (say light years away) from the body at rest. When our body
is allowed to rotate then the potential $\phi$  determining field angular momentum rotates together with it. Hence, in spite of the huge distance, the amount of this quantity at the far ends of the solenoid {\em varies synchronously} with the rotation of the body without any delay. Though from a purely computational point of view this apparent nonlocal behaviour presents no difficulty a simple explanation of it in terms of local interactions alone would certainly improve our understanding of this important aspect of electrodynamics\footnote{Another problem related to the nonlocal behaviour of the Coulomb gauge potential is the finiteness of signal (light) velocity. A substantial amount of literature has been devoted to this subject since W. Heisenberg proposed it to his disciple 
S. Kikuchi [2] and E. Fermi published his study of quantum electrodynamics [3]. A recent discussion of this problem in a variety of gauges is found in [4] where references to earlier literature are also found.}. For this reason in the next section the results arrived at above will be rederived by means of explicitely local considerations.

\section{The underlying physical picture}

Consider now a point charge moving arbitrarily along the trajectory ${\bf x} = {\bf r}(t)$ in a stationary magnetic field ${\bf B}({\bf x})$  {\em of finite extension}. The Hamiltonian is
\[H = \frac{1}{2m}\left ({\bf p} - \frac{e}{c}{\bf A}({\bf r})\right )^2.\]
Since the Lorentz-force acting on the charge does not perform any work on it this $H$ is numerically equal to the kinetic energy $mV^2/2$. Hence ${\bf p} = m{\bf V} + (e/c){\bf A}({\bf r})$ where ${\bf V} = \dot{\bf r}(t)$ is the instantaneous velocity of the charge.

Any gauge transformation of $\bf A$ is accompanied by a canonical transformation of $\bf p$ so as to make the difference ${\bf p}-(e/c){\bf A}$ unchanged. As it is well known (see e.g. [5]) in the special case of the Coulomb-gauge the quantity $(e/c){\bf A}$ is equal to the momentum $\bf G$ stored in the combined electric field of the charge and the external field $\bf B$ when the charge is {\em at rest} at the point ${\bf r}(t)$ of its trajectory. Hence, this $\bf G$ seems rigidly attached to the particle as was the case with the field angular momentum $\bf N$ and the rotator in the example of the previous sections. At any moment of time $t$ and point $P$ in space the electric field contributing to $\bf G$ is determined by the location of the charge ${\bf r}(t)$ at the same moment of time however far away from $P$, and, therefore, {\em it obviously cannot be equal to the true momentum $\pmb{\cal G}$ stored in the combined field of the particle and that of the external sources at} $t$. Since composite charged bodies are built up of pointlike charged constituents the nonlocality found in the case of the rotating body has been inherited from this description of the point charge.

The sole virtue of the example with the rotator in the previous section was that, owing to the splitting of $\bf N$ into fractions located within disjoint domains of space, its nonlocality appeared more spectacular than that of a point charge. But in order to elucidate the physical picture behind this seemingly nonlocal description it is sufficient to confine ourselves to the case of a single point charge.

The electromagnetic part of the energy-momentum tensor decomposes into the sum of three terms $T_{\alpha\beta}^{(i)}$ according to the power of $\bf B$ in them. Each term satisfies a balance equation (continuity equation with sources) of its own. This may be verified either by direct computation or by rescaling the external field $\bf B$ with a scalar factor $k$. For an arbitrary value of $k$ the electromagnetic energy-momentum tensor becomes equal to the sum $k^0T_{\alpha\beta}^{(0)} + k^1T_{\alpha\beta}^{(1)} + k^2T_{\alpha\beta}^{(2)}$. Since the sum satisfies a balance equation for any value of $k$ the three parts must satify it separately.

The term of second order, being constant in time, is obviously of no interest. The term independent of the external field describes radiation and radiation reaction (self-interaction) of the charge and obeys a balance equation with a source equal to the negative of the self-force. This term is of paramount importance of its own right but for the external field problems only the mixed part $T_{\alpha\beta}^{(1)}$ of the tensor is of significance. The sources of the balance equation satisfied by this term in an external magnetic field are the negative of the Lorentz-force acting on the charge in the external field and the minus Lorentz-force density experienced by the sources of this field from the side of the moving charge. From our point of view a cardinal property of this part of the energy-momentum tensor is the absence of radiation. This follows from the finite extension of the external field in space.

The force experienced by the charge at moment $t_0$ does not depend on its subsequent motion. Therefore, in calculating this force we can rely on the {\em truncated} (at $t=t_0$){\em trajectory} which is obtained from the true trajectory by bringing the charge to a stop at the position ${\bf r}_0 = {\bf r}(t_0)$. At this moment the true spatial distribution of the momentum density is certainly different from the momentum density $(1/4\pi c)[\En ,\,{\bf B}]$ of a charge in state of rest at ${\bf r}_0$ but the total field momentum will tend to the integral
\begin{equation}
\Gn = \frac{1}{4\pi c}\int d^3x\,[\En ,\,{\bf B}]\labl{B2a}
\end{equation}
as $t\longrightarrow\infty$ \footnote{The upper index zero on a quantity indicates that it refers to the truncated motion.}. The reason is that due to the absence of radiation the mixed field momentum emitted by the moving charge before $t_0$ will be completely absorbed  in later times by the sources of the external field.

Let us compare momentum exchange between the particle and the field in the two motions truncated at moments of time $t_0$ and $t_0 + dt_0$. The field momentum emitted by the particle in the infinitesimal interval $(t_0 < t < t_0 + dt_0)$ consists of two parts: that what remains stored in the form of field momentum forever and an additional part which will be subsequently absorbed by the sources of the external field. Accordingly, the force experienced by the particle consists of two contributions:
\begin{equation}
\Fn = - \frac{d\Gn}{dt_0} - \Fnp ,\labl{B2b}
\end{equation}
where \Fnp\ is the Lorentz-force experienced by the sources of $\bf B$ when some portion of the mixed field momentum emitted at $t_0$ by the charge is absorbed.

Since \Gn\ depends on time through the motion of the charge alone we have
\begin{equation}
\frac{d\Gnc{\alpha}}{dt_0} = V_\beta(t_0)\partial_\beta\Gnc{\alpha} = ,\labl{B3a}
\end{equation}
where ${\pmb{\partial}}$ denotes derivation with respect to $\bf r$.

The force \Fnp\ is related to the momentum difference received by the sources of $\bf B$ in the motions truncated at the moments $t_0 + dt_0$ and $t_0$. Hence, in order to evaluate this force we must first calculate the momentum \Gnp\ received by the sources via the retarded field of the charge. It is equal to the time integral of the Lorentz-force experienced by them:
\begin{equation}
\Gnp = \frac{1}{c}\int dt\cdot d^3x\,[{\bf J}({\bf x}),\,\Bnp ({\bf x},t)],\labl{B3b}
\end{equation}
where ${\bf J}({\bf x})$ is the current density supporting the external magnetic field. The retarded magnetic field of the charged particle is
\[\Bnp ({\bf x},t) = \Rot\Anp ({\bf x},t).\]
The vector potential here is the Li\'{e}nard-Wiechert potential of the particle's current density ${\bf j}^{(0)}$:
\begin{equation}
\Anp ({\bf x},t) = \frac{1}{c}\int dt'\cdot d^3x'\,\frac{\delta(t'-t+|{\bf x} - {\bf x}'|/c)}{|{\bf x} - {\bf x}'|}{\bf j}^{(0)}({\bf x}',t').\labl{B3d}
\end{equation}
Since at $t = t_0$ the particle comes to a stop we can write
\[{\bf j}^{(0)}({\bf x}',t') = {\bf j}({\bf x}',t')\Theta (t_0 - t'),\]
where $\Theta (t)$ is the step-function. Then 
\[\int_{-\infty}^{\infty}dt\, \Anp ({\bf x},t) = \frac{1}{c}\int\frac{d^3x'}{|{\bf x} - {\bf x}'|}\int_{-\infty}^{t_0}dt'\,{\bf j}({\bf x}',t').\]
Now, using the relation
\begin{equation}
\Rot(f{\bf U}) = [\Grad f,\,{\bf U}] + f\Rot{\bf U}\labl{B4b}
\end{equation}
we calculate the curl of this function:
\begin{eqnarray*}
\int_{-\infty}^{\infty}dt\, \Bnp ({\bf x},t) = \Rot \int_{-\infty}^{\infty}dt\, \Anp ({\bf x},t) =\\ 
= \frac{1}{c}\int d^3x'\,\left [\Grad|{\bf x} - {\bf x}'|^{-1},\,\int_{-\infty}^{t_0}dt'\,{\bf j}({\bf x}',t')\right ].
\end{eqnarray*}
To obtain \Fnp\ we have to compute the derivative of \Gnp\ as given in \reff{B3b} with respect to $t_0$. At first we write
\[\frac{d}{dt_0} \int_{-\infty}^{\infty}dt\, \Bnp ({\bf x},t) = \frac{1}{c}\int d^3x'\,\left [\Grad|{\bf x} - {\bf x}'|^{-1},\,{\bf j}({\bf x}',t_0)\right ].\]
The current density of a pointlike charged particle is
\[{\bf j}({\bf x}',t_0) = e{\bf V}(t_0)\delta \left ({\bf x}' - {\bf r}(t_0)\right ).\]
Then
\begin{equation}
\frac{d}{dt_0} \int_{-\infty}^{\infty}dt\, \Bnp ({\bf x},t)  = \frac{e}{c}\left [\Grad |{\bf x}-{\bf r}(t_0)|^{-1},\,{\bf V}(t_0)\right ] = \frac{1}{c}[\Grad\phi^{(0)}, {\bf V}(t_0)],\labl{B4a}
\end{equation}
where
\[\phi^{(0)}({\bf x}) = \frac{e}{|{\bf x}-{\bf r}(t_0)|}\]
is the potential of a charge at ${\bf x} = {\bf r}(t_0)$.

Now we can write
\begin{eqnarray*}
\Fnp = \frac{d\Gnp}{dt_0} = \frac{1}{c^2}\int d^3x \left [{\bf J}({\bf x}),\,[\Grad\phi^{(0)},\,{\bf V}(t_0)]\right ] = \\
= \frac{1}{c^2}\int d^3x \left \{\left ({\bf J}\cdot{\bf V}(t_0)\right )\Grad\phi^{(0)} - ({\bf J}\cdot\Grad\phi^{(0)}){\bf V}(t_0)\right \}
\end{eqnarray*}
The second integral on the r.h.s. is equal to zero because the stationary current ${\bf J}({\bf x})$ is solenoidal:
\[\int d^3x\,{\bf J}\cdot\Grad\phi^{(0)} = - \int d^3x\, \phi^{(0)}\,\Div{\bf J} = 0.\]
In Cartesian coordinates we have
\[\nabla_\alpha\phi^{(0)} \equiv \frac{\partial\phi^{(0)}}{\partial x_\alpha} =
-\frac{\partial\phi^{(0)}}{\partial r_\alpha} \equiv - \partial_\alpha\phi^{(0)}.\]
Then
\begin{equation}
\Fnp = -\frac{1}{c^2}\partial_\alpha\int d^3x\,\left ({\bf J}\cdot{\bf V}(t_0)\right )\phi^{(0)} = 
- V_\beta(t_0)\partial_\alpha\left [\frac{1}{c^2}\int d^3x\, \phi^{(0)}J_\beta\right ].\labl{B5a}
\end{equation}
Equation $\Rot{\bf B} = (4\pi /c){\bf J}$ permits us to replace  $\bf J$ by $\bf B$:
\[\frac{1}{c^2}\int d^3x\, \phi^{(0)}J_\beta = \frac{1}{4\pi c}\int d^3x\,\phi^{(0)}\Rot{\bf B}.\]
Here we can use \reff{B4b} again:
\[\frac{1}{c^2}\int d^3x\, \phi^{(0)}J_\beta = \frac{1}{4\pi c}\left\{\int d^3x\,[{\bf B},\,\Grad\phi^{(0)}] +
\int d^3x\,\Rot (\phi^{(0)}{\bf B})\right\}\]
The second term is transformed into a surface integral and can be dropped. In the first term $\Grad\phi^{(0)} = - \En$. Then, using \reff{B2a}, we obtain
\[\frac{1}{c^2}\int d^3x\, \phi^{(0)}J_\beta = \frac{1}{4\pi c}\int d^3x\,[\En ,\, {\bf B}] = \Gn .\]
Substitution of this expression into \reff{B5a} leads to the concise form
\begin{equation}
\Fnpc{\alpha} = -V_\beta (t_0)\partial_\alpha\Gnc{\beta}.\labl{B6b}
\end{equation}
From now on we may drop upper 0 indices. Inserting \reff{B6b} and \reff{B3a} into \reff{B2b} we obtain for the force acting on the charge the formula
\begin{equation*}
F_\alpha = V_\beta\partial_\alpha G_\beta - V_\beta\partial_\beta G_\alpha ,
\end{equation*}
where ${\bf V}(t) = \dot{\bf r}(t)$ and ${\bf G}$ is equal to the field momentum stored in the combined field of a charge {\em resting at} ${\bf r}(t)$. Since $\bf V$ is independent of $\bf r$, the vector form of this equation is
\begin{equation}
{\bf F} = \left [{\bf V},\,\Rot {\bf G}\right ] =
\bigl ({\pmb\nabla}({\bf V}\cdot{\bf G}) - ({\bf V}\cdot {\pmb\nabla }){\bf G}\bigr ),\labl{B6a}
\end{equation}
where $\pmb{\nabla}$ denotes now differentiation with respect to $\bf r$ the only coordinate $\bf G$ depends on.

On the other hand, we know that $\bf F$ is the Lorentz-force which acts on the charge:
\[{\bf F} = \frac{e}{c}[{\bf V},\,{\bf B}] = \left [{\bf V},\,\Rot \frac{e}{c}{\bf A}\right ].\]
We see that (up to the factor $e/c$) $\bf G$ can be identified with the unique Coulomb gauge vector potential vanishing at infinity together with $\bf G$. Equation \reff{B6a} can, therefore, be written also as
\begin{equation}
{\bf F} = \frac{e}{c}\bigl ({\pmb\nabla}({\bf V}\cdot{\bf A}) - ({\bf V}\cdot {\pmb\nabla }){\bf A}\bigr ).\labl{B6aa}
\end{equation}

At the beginning of the present section we have stressed the apparent nonlocality of this gauge. Now we see explicitely that locality is by no means broken: There is in fact no field momentum attached rigidly to the charge and in the general case the true mixed field momentum $\pmb{\cal G}$ differs from $\bf G$ at any moment of time\footnote{Analogously, the quantity $\bf N$ in \reff{6b} is in fact different from the true mixed field angular momentum $\pmb{\cal N}$ at any moment of time.}. The flow of the combined field momentum density obeys continuity equation in space between the moving charge and the sources of the external field and this is what locality means. 

\section{Splitting of the Lorentz force}

A bonus of the preceding investigation is the recognition that in the external field problem the Lorentz force decomposes into two terms each with clear physical meanings. When in \reff{B6aa} $\bf A$ is that unique Coulomb gauge vector potential which vanishes at infinity than the second term describes momentum exchange of the  charge with the field while the first one corresponds to the charge's momentum exchange (via its retarded field) with the currents and magnets supporting the external magnetic field\footnote{When the external field is electric, no mixed field momentum arises. It is for this reason that the Coulomb force $-e{\pmb\nabla}\Phi$ is of purely first type.}. Naturally, this decomposition is gauge invariant. 

An interesting special example is that of a charged particle passing by an infinitely long straight ideal solenoid (a classical `Aharonov-Bohm situation'). Since the magnetic field outside the solenoid is zero the two terms in \reff{B6aa} cancel each other but neither of them vanishes. That means that {\em the charge `catalizes' momentum exchange between the combined field and the coil}. Outside the solenoid we have
\begin{equation}
{\bf A} = \frac{S}{2\pi}\cdot\frac{[{\bf B},\,{\bf r}]}{r^2},\labl{outside}
\end{equation}
where $S$ is the cross section of the solenoid. In this domain the r. h. s. of \reff{B6a} are equal to zero and so \reff{B6b} can be written as $F'_\alpha = - V_\beta\partial_\beta G_\alpha$. Since ${\bf G} = (e/c){\bf A}$, the total momentum imparted to the coil by the moment $t$ when the charge reaches position $\bf r$ is equal to
\begin{equation}
{\bf p'} = \int_{-\infty}^t dt\,{\bf F}'  = - \frac{eS}{2\pi c}\cdot\frac{[{\bf B},\,{\bf r}]}{r^2}.\labl{transf}
\end{equation}
When the charge leaves freely to infinity no net momentum exchange occurs. But if it is stopped at $\bf r$ a finite amount \reff{transf} of momentum is transferred from the combined field to the coil. I wonder whether this effect may be experimentally observed.

Inside the solenoid the continuation of the external vector potential \reff{outside} is given by \reff{2a} and the field there is homogeneous. Inserting this into \reff{B6aa} we find that, for a point charge revolving in a plane perpendicular to solenoid's symmetry axis on a circle around it, the two terms on the r. h. s. are equal to each other. Therefore, the reaction to the centripetal force acting on the charge is distributed evenly between the field and the coil: the force experienced by the coil is the half of the force experienced by the revolving point charge.

\section{Relation to the Aharonov-Bohm effect}

When an electron passes by the solenoid no net momentum is transferred to the latter. But the {\em action} of the coil, considered as the target in the scattering with the electron, changes by a finite amount.

Choose the axis of the solenoid as the $z$-direction and assume that the trajectory of the electron lies in the $xy$ plane at $y = \pm a$. Then the change of the coil's action is
\begin{equation}
{\cal S}_{\pm}' = \int_{-\infty}^\infty p_x'dx = \pm\frac{eSBa}{2\pi c}\int_{-\infty}^\infty\frac{dx}{r^2} = \pm\frac{eSB}{2c}.\labl{Sprime}
\end{equation}

In quantum mechanics this peculiar elastic scattering process can be described in quasiclassical approximation considering both the electron and the coil as quantum systems. The  initial wave-function is $\psi\Psi$. The first factor refers to the electron, the second to the coil. Then the final wave function  (the out state) is given by the superposition
\begin{equation}
\psi_+(\mathbf r)\Psi_+ + \psi_-(\mathbf r)\Psi_-.\labl{wf1}
\end{equation}
The quasiclassical wave functions $\psi_\pm$ for the electron describe its motion in the vicinity of the classical trajectories $y = \pm a$ respectively, while the wave functions $\Psi_\pm$ of the coil are  $\di e^{i{\cal S}_\pm' /\hslash}\Psi$. Then \reff{wf1} becomes
\begin{equation}
\left (\psi_+(\mathbf r)e^{i{\cal S}_+' /\hslash} + \psi_-(\mathbf r)e^{i{\cal S}_-' /\hslash}\right )\Psi .\labl{wf2}
\end{equation}
{\em The relativ phase} between the two terms of \reff{wf1} contains obviously the well-known {\bf B}-dependent Aharonov-Bohm phase 
\[({\cal S}_+' - {\cal S}_-')/\hslash = \frac{eSB}{\hslash c}.\]

It is, perhaps, worth pointing out that the reasoning which, beginning with section 3, led us finally  to this formula was based entirely on the fields {\bf E} and {\bf B}. The Coulomb-gauge vector potential {\bf A} appeared merely as a {\em notation} for the field momentum $\bf G^{(0)}$ (divided by $e/c$). 

\appendix
\section*{Appendix. Derivation of eq. \reff{4a}}

Let ${\bf N}$ denote the volume integral in \reff{3b} confined to a sphere of radius $R$. Inserting ${\bf E} = -\Grad\phi$ and ${\bf B} = \Rot{\bf A}$ into it we obtain after some rearragements
\[N_\alpha = -\frac{1}{4\pi c}\epsilon_{\alpha\beta\gamma}\int_R d^3x\; x_\beta\left\{\nabla_\delta\phi\cdot\nabla_\gamma A_\delta - \nabla_\delta\phi\cdot\nabla_\delta A_\gamma\right\},\]
where $\epsilon_{\alpha\beta\gamma}$ is the Levi-Civita symbol.

In the second term the two $\nabla$ symbols have identical indices. By partial integration we make them to merge into $\nabla^2$:
\[N_\alpha = -\frac{1}{4\pi c}\epsilon_{\alpha\beta\gamma}\int_R d^3x\; \left\{x_\beta\nabla_\delta\phi\cdot\nabla_\gamma A_\delta - \nabla_\delta(x_\beta\cdot\nabla_\delta\phi\cdot A_\gamma) +\nabla_\beta\phi\cdot A_\gamma + x_\beta\cdot\nabla^2\phi\cdot A_\gamma\right\}.\]

In the first term the index of the first $\nabla$-symbol and that of the vector potential are identical. By further partial integration we form $\Div\bf A$ out of them:
\begin{eqnarray*}
N_\alpha = -\frac{1}{4\pi c}\epsilon_{\alpha\beta\gamma}\int_R d^3x\; \left\{\nabla_\delta (x_\beta\cdot\phi\cdot\nabla_\gamma A_\delta )- \nabla_\gamma (\phi A_\beta ) + \nabla_\gamma\phi\cdot A_\beta - \right.\\
\left. - \phi\cdot x_\beta\nabla_\gamma(\Div {\bf A})- \nabla_\delta (x_\beta\cdot\nabla_\delta\phi\cdot A_\gamma) + \nabla_\beta\phi\cdot A_\gamma + x_\beta\cdot\nabla^2\phi\cdot A_\gamma\right\}
\end{eqnarray*}
By eq. \reff{2a} we have $\Div{\bf A} = 0$. Out of the six terms left we have three pure divergences (surface terms). Two of the three volume terms cancel each other for symmetry reason:
\[\epsilon_{\alpha\beta\gamma} (\nabla_\gamma\phi\cdot A_\beta + \nabla_\beta\phi\cdot A_\gamma) = 0.\]
If we insert $\nabla^2\phi = -4\pi\rho$ into the remaining one we obtain the r.h.s. of \reff{1b} the integration being confined to the sphere of radius $R$:
\[{\bf N}^{(0)} = +\int_R d^3x\,\left [{\bf x}, \frac{\rho}{c}{\bf A}\right ].\]
According to \reff{3a}, its value is equal to
\[N_\alpha^{(0)} = -\frac{1}{6c}D_{\alpha\beta}B_\beta + \frac{1}{3c}\langle\rho r^2\rangle B_\alpha\]
for any sphere outside the body.

The three surface integrals are
\begin{align*}
N_\alpha^{(1)} & = +\frac{1}{4\pi c}\epsilon_{\alpha\beta\gamma}\int_R d^3x\;\nabla_\gamma(\phi A_\beta )\\
N_\alpha^{(2)} & = -\frac{1}{4\pi c}\epsilon_{\alpha\beta\gamma}\int_R d^3x\;\nabla_\delta (x_\beta\phi\cdot\nabla_\gamma A_\delta )\\
N_\alpha^{(3)} & = +\frac{1}{4\pi c}\epsilon_{\alpha\beta\gamma}\int_R d^3x\;\nabla_\delta (x_\beta\cdot\nabla_\delta\phi\cdot A_\gamma ).
\end{align*}
All of them must be of the form $(\eta_i /c)D_{\alpha\beta}B_\beta$ ($i = 1,2,3$) and our task is to calculate the coefficients $\eta_i$.

Using Gauss theorem 
\[\int_R d^3x\,\nabla_\alpha f = \oint_R dS_\alpha\cdot f = \oint R^2d\Omega\; \frac{x_\alpha}{R}f,\]
where $\Omega$ denotes solid angle, we have
\begin{equation}
N_\alpha^{(1)} = +\frac{1}{c}\epsilon_{\alpha\beta\gamma}\frac{R}{4\pi}\oint d\Omega\; x_\gamma\phi A_\beta .\tag{A.1}
\end{equation}
The body is uncharged by hypotheses and so no Coulomb potential contributes to $\phi$. The contribution of its dipole moment $\bf d$ to ${\bf N}^{(1)}$ is of the form $constant\times [{\bf d},\, {\bf B}]$ but the constant must be zero because it is a polar vector while ${\bf N}^{(1)}$ is an axial one. From moments higher than quadrupole no vector can be formed with $\bf B$. Hence it is only the quadrupole potential
\begin{equation}
\phi = \frac{1}{2R^5}x_\alpha x_\beta D_{\alpha\beta}\tag{A.2}
\end{equation}
which contributes to the surface integrals. 

For the vector potential we can substitute
\[A_\beta = -\frac{1}{2}\epsilon_{\beta\rho\sigma}x_\rho B_\sigma ,\]
and we obtain
\[N_\alpha^{(1)} = -\frac{1}{2c}\epsilon_{\alpha\beta\gamma}\epsilon_{\beta\rho\sigma}B_\sigma\frac{R}{4\pi}\int d\Omega\;x_\gamma x_\rho\phi = -\frac{R}{8\pi c}\left\{B_\alpha R^2\int d\Omega\;\phi + B_\gamma\int d\Omega\;x_\alpha x_\gamma\phi\right\} .\]
The first integral vanishes because of the zero net charge condition.
Substituting (A.2) into the second term and using the relation
\[\int d\Omega\; x_\gamma x_\mu x_\nu x_\alpha = \frac{4\pi R^4}{15}(\delta_{\gamma\mu}\delta_{\nu\alpha} + \delta_{\gamma\nu}\delta_{\mu\alpha} + \delta_{\gamma\alpha}\delta_{\mu\nu})\]
we arrive at the value  $\eta_1 = 1/30$.

A completely analogous derivation leads to $\eta_2 = -1/30$. Hence, the first two surface contributions cancel each other.

The remaining surface integral is
\[N_\alpha^{(3)} = +\frac{1}{c}\epsilon_{\alpha\beta\gamma}\frac{R}{4\pi}\int d\Omega\; x_\beta x_\delta\nabla_\delta\phi\cdot A_\gamma .\]
Since the potential (A.2) is of degree $-3$ in $r$ we have
\[x_\delta\nabla_\delta\phi =  r\frac{\partial\phi}{\partial r} = -3\phi .\]
Hence
\[N_\alpha^{(3)} = -\frac{3}{c}\epsilon_{\alpha\beta\gamma}\frac{R}{4\pi}\int d\Omega\; x_\beta A_\gamma .\]
Comparing this with (A.1) we obtain $\eta_3 = +1/10$. Adding together ${\bf N}^{(0)}$ and the surface contributions \reff{4a} is obtained.

When the body is situated on the axis of an infinitely long straight ideal solenoid of radius $R_s$ than \reff{4a} remains valid for the field angular momentum within the sphere provided $R \le R_s$. The total field angular momentum, however, is given by \reff{3a} (i.e. ${\bf N}^{(0)}$) because the surface integrals vanish as $R\longrightarrow\infty$. Outside the solenoid the vector potential is given by \reff{outside} rather than \reff{2a} and tends to zero as $R$ increases. Within the solenoid \reff{2a} remains true at arbitrarily large distances from the body but the surface of the parts of the sphere remain finite and the integral vanishes on them for this reason.

\section*{References}
\begin{description}
\item{[1]} J. D. Jackson, Classical Electrodynamics, 3rd ed. (Wiley, New York, 1999).  
         
\item{[2]} S. Kikuchi, Zs.f.Phys. {\bf 66}, 558, (1930)\\
	http://www.springerlink.com/content/v96267q8182ru807/

\item{[3]} E. Fermi, Rev. Mod. Phys. {\bf 4}, 87, (1932)\\
	http://link.aps.org/doi/10.1103/RevModPhys.4.87  
	
\item{[4]} J. D. Jackson,  Am. J. Phys. {\bf 70}, 917 (2002)\\	http://arxiv.org/abs/physics/0204034
	
\item{[5]} N. Kroll in {\it Quantum Optics and Electronics, Lectures delivered at Les Houches 1964} Gordon and Breach 1965
\end{description}
\end{document}